# Demonstration of inkjet-printed targets for activation cross-section measurements: deuteron-induced reactions on strontium up to 24 MeV

Akihiro Nambu [a,b,*], Masayuki Aikawa [a,b,c,d], Yudai Shigekawa [a], Yousuke Kanayama [a], Tomohiro Tomitsuka [a], Sayantani Mitra [a], and Hiromitsu Haba [a]

[a] *Nishina Center for Accelerator-Based Science, RIKEN, Wako 351-0198, Japan*

[b] *Graduate School of Biomedical Science and Engineering, Hokkaido University, Sapporo 060-8638, Japan*

[c] *Faculty of Science, Hokkaido University, Sapporo 060-0810, Japan*

[d] *Global Center for Biomedical Science and Engineering, Faculty of Medicine, Hokkaido University, Sapporo 060-8648, Japan*

Abstract

Excitation functions for the $^{nat}$Sr($d,x$) reactions producing $^{86m,g}$Y, $^{87m,g}$Y, and $^{88}$Y were measured up to 24-MeV deuteron energy using the stacked-foil technique and gamma-ray spectrometry. The $^{nat}$Sr targets were prepared using an inkjet printing method to ensure sufficient uniformity. To the best of our knowledge, this is the first experimental report of excitation functions measured using inkjet-printed targets, demonstrating the applicability of this technique to cross-section measurements. The measured cross sections were discussed in comparison with literature data and TENDL-2023 predictions. Our cross sections were generally consistent with those reported in the previous study, but were slightly higher in the threshold energy region. The TENDL-2023 library generally overestimates the experimental values. We report the experimental cross sections for the $^{nat}$Sr($d,x$)$^{86m}$Y reaction for the first time. These measured cross sections contribute to the refinement of nuclear reaction databases for practical applications.

## Introduction

Yttrium radioisotopes play a pivotal role not only in fundamental fields such as physics and chemistry, but also in applied fields such as engineering and medicine. $^{90}$Y ($T_{1/2}$ = 64.0 h, $\beta^{-}$=100%) is a well-established nuclide widely used for internal radiotherapy, including microsphere brachytherapy and radioimmunotherapy [1,2]. In the context of theranostics, $^{86g}$Y ($T_{1/2}$ = 14.7 h, EC = 72.8%, $\beta^+$=27.2%) has attracted significant attention as a positron-emitting surrogate for $^{90}$Y [3]: its $\beta^+$ emission allows for quantitative biodistribution imaging and dosimetry via Positron Emission Tomography (PET). $^{87g}$Y ($T_{1/2}$ = 79.8 h, EC = 99.82%, $\beta^+$ = 0.18%) serves as a γ-ray emitter for tracing biodistribution[4] and as the parent nuclide for the $^{87g}$Y/$^{87m}$Sr generator system, which has been proposed for bone scintigraphy [5,6]. $^{88}$Y ($T_{1/2}$ = 106.6 d, EC = 99.79%, $\beta^+$=0.21%) serves as a long-lived reference source for γ-ray detectors due to its characteristic high-energy γ-ray emissions [7].

* Corresponding author: Akihiro NAMBU (akihiro.nambu@riken.jp), Nishina Center for Accelerator-Based Science, RIKEN, Wako 351-0198, Japan

Charged-particle induced reactions on strontium targets serve as promising production routes for yttrium isotopes, specifically $^{86g}$Y, $^{87g}$Y, and $^{88}$Y. For proton-induced reactions, excitation functions on enriched $^{86}$Sr, $^{88}$Sr, and natural strontium ($^{nat}$Sr) targets have been systematically measured, enabling reliable evaluations of production yields and radionuclidic purity [8–25]. In contrast, experimental data for deuteron-induced reactions remain scarce. While cross sections for $^{86}$Sr and $^{nat}$Sr targets have recently been reported [26,27], these remain the only available data, according to a survey using the EXFOR library[28]. Furthermore, data for the metastable state $^{86m}$Y, which affects the radionuclidic purity of the ground state, are lacking. Considering these backgrounds, further experiments and measurements are obviously essential.

One of the primary challenges in measuring excitation functions for strontium is fabricating thin and uniform targets, because metallic strontium is highly chemically reactive. In recent studies, strontium carbonate targets prepared via sedimentation have been used [23–25,29]. However, the sedimentation is a highly labor-intensive target fabrication technique for achieving sufficient target uniformity and stability, requiring optimization of the amounts of target material, ethyl cellulose, and acetone solvent [29].

To address these issue, we adopted a novel target fabrication technology using a commercial inkjet printer [30], which is expected to provide high uniformity of target thickness. In this method, target material dissolved in an aqueous solution is deposited onto a backing foil as nanoliter-sized droplets. Previous studies have demonstrated that uniform targets can be achieved for lanthanum, gold, and vanadium [30,31]. To the best of our knowledge, no excitation functions have been measured using inkjet-printed targets to date; however, this technology is promising and applicable to a wide range of target nuclides in various chemical forms.

In the present study, we fabricated the $^{nat}$Sr targets using the inkjet-printing method and demonstrated that they are applicable for measuring the excitation functions for $^{86m,86g}$Y, $^{87m,87g}$Y, and $^{88}$Y produced via deuteron-induced reactions on $^{nat}$Sr up to 24 MeV. The measured cross sections were compared with the literature data[27] and the theoretical predictions from the TENDL-2023 library [32].

## Method

Cross sections for the deuteron-induced reactions on $^{nat}$Sr were determined using the stacked-foil activation technique combined with γ-ray spectrometry. The $^{nat}$Sr targets were fabricated from strontium nitrate ($^{nat}Sr(NO_3)_2$) solution using a commercial inkjet printer (PJK-200S, MICROJET Corporation). The chemical procedure for preparing the $^{nat}Sr(NO_3)_2$ solution is presented in Fig. A.1 in Appendix A. No-carrier-added $^{85}$Sr, produced in the $^{nat}$Rb($d$,$xn$)$^{85}$Sr reaction, was added to the $^{nat}Sr(NO_3)_2$ solution to quantify the amount of deposited $^{nat}$Sr. The two batches of the $^{nat}Sr(NO_3)_2$ solution (#1: $Sr^{2+}$ 11.43 mg/mL, $^{85}$Sr 8.75 kBq/mL, and specific radioactivity 765 Bq/mg; #2: $Sr^{2+}$ 11.48 mg/mL, $^{85}$Sr 5.16 kBq/mL, and specific radioactivity 449 Bq/mg; 0.01 M nitric acid solution) were prepared.

The $^{nat}Sr(NO_3)_2$ solution was deposited onto 10-µm-thick aluminum backing foils (purity: 99.999%, thickness: 2.75 mg/cm$^2$ or 2.61 mg/cm$^2$; Goodfellow Co., Ltd., UK) heated on a hot plate at 70°C. According to the reference [33], the chemical composition of deposit $^{nat}$Sr targets is estimated to be $^{nat}Sr(NO_3)_2$ crystal without hydration. The printing process involved depositing a total of 464 droplets (approx. 2.9 nL each) with a pitch of 0.333 mm, distributed uniformly over a defined area of 0.5145 cm$^2$. The target thickness was determined by measuring the 514.0-keV γ-ray (intensity $I_\gamma$ = 96%) of the $^{85}$Sr tracer using a germanium detector, yielding an average thickness of approximately 70 and 90 µg/cm$^2$ as $^{nat}$Sr for the two prepared batches.

The spatial uniformity of the targets was evaluated by autoradiography. The surfaces of target foils were covered with a Mylar film (1.5 µm, TORAY INDUSTRIES, INC.) to prevent contamination. An imaging plate (Amersham BAS-IP SR 2040 E) was placed over the foils and kept optically shielded for 58 hours. The data was acquired at a spatial resolution of 50 µm using Amersham TYPHOON scanner RGB. The intensity of photostimulated luminescence (PSL) was saved as an 8-bit (256-level) grayscale bitmap image. According to the simulation using the PHITS 3.24 code [34], conversion electrons are the primary contributors to PSL. Autoradiographic analysis revealed that the uncertainty in the target thickness is within 3% (see Appendix B). Figures 1(a) and 1(b) show typical images obtained by the optical scanner (CanoScan LiDE 400, Canon Marketing Japan Inc.) and the autoradiographic scanner, respectively. Finally, each target was covered with a 10-µm-thick $^{27}$Al foil (purity: 99.999%, thickness: 2.75 mg/cm$^2$; Goodfellow Co., Ltd., UK) to protect the deposited $^{nat}Sr(NO_3)_2$ and to collect recoil products from the target.

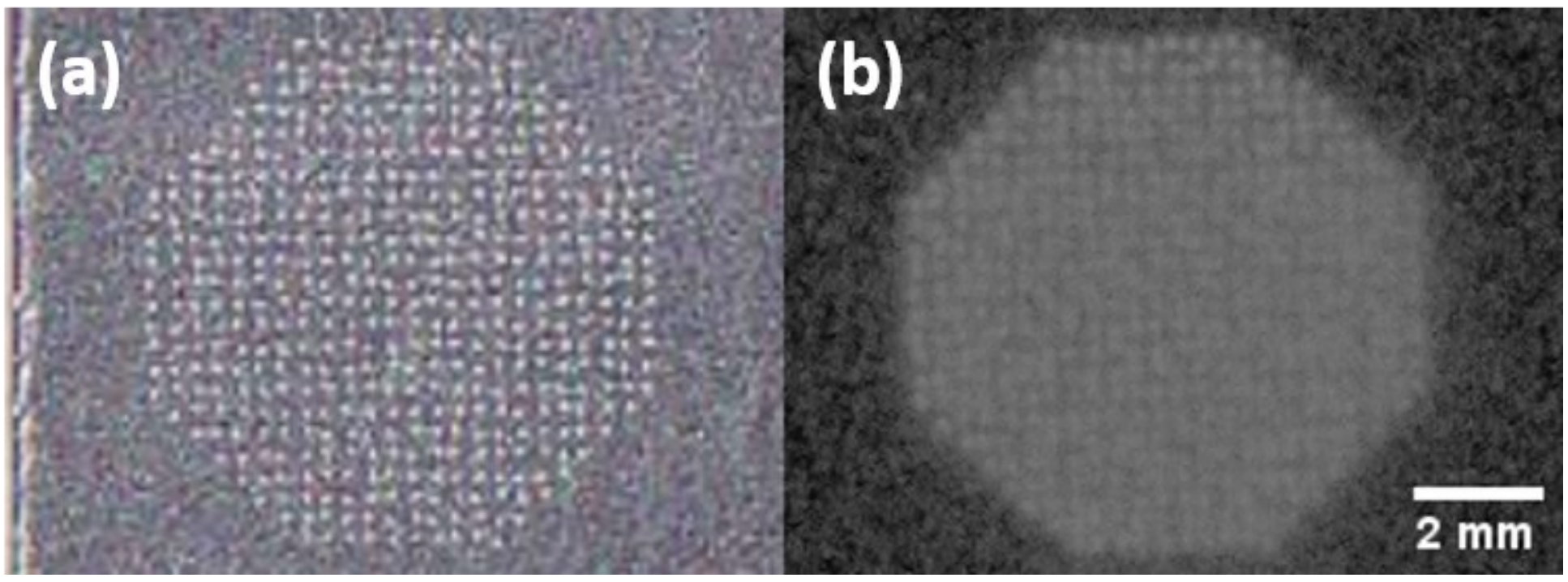

Fig. 1. Typical images of the $^{nat}Sr(NO_3)_2$ target taken by the optical scanner (a) and by the autoradiographic scanner (b).

The stacked-foil target assembly was constructed by interleaving the inkjet-printed $^{nat}Sr(NO_3)_2$ targets, $^{nat}Cu$ beam monitor foils (purity: 99.99%, thickness: 21.7 mg/cm$^2$; Nilaco Corp., Japan), and $^{27}Al$ recoil catchers (purity: 99.999%, thickness: 2.75 mg/cm$^2$ and 2.61 mg/cm$^2$; Goodfellow Co., Ltd., UK). Specifically, 15 sets of target stacks, configured as $^{27}Al$ (cover)-$^{nat}Sr(NO_3)_2$-$^{27}Al$ (backing), and 16 sets of monitor stacks, configured as $^{27}Al$-$^{nat}Cu$-$^{27}Al$, were stacked alternately. The precise thickness of each metallic foil was determined gravimetrically. Table 1 summarizes the target specifications.
The irradiation was performed at the AVF cyclotron of the RIKEN RI Beam Factory, Wako, Japan. The incident deuteron energy was determined to be 24.5±0.1 MeV via the time-of-flight (TOF) method [35]. The incident beam was collimated by $\phi$3-mm aperture placed forward the target. Each beam intensity at the target and at the collimator was individually monitored using a Faraday cup connected to a current integrator. The beam intensity at the target was maintained at 99.3 nA in average for 30 min. The beam parameters are also summarized in Table 1.

Energy degradation along the stack was calculated using stopping powers derived from the SRIM-2013 code [36]. The uncertainty in the projectile energy was estimated by considering the cumulative energy straggling within the target stack arising from uncertainties in the initial beam energy and the target foil thicknesses. Additionally, the inkjet-printed dot targets cause spatial variations in energy loss, leading to a further broadening of the beam energy distribution. This effect was quantitatively evaluated via Monte Carlo simulations and incorporated into the energy uncertainty. In the simulation, the dot target was modeled based on observations of the strontium target using a scanning electron microscopy (SEM) (TM4000 Plus, Hitachi High-Tech Corporation). Figure 2 displays the SEM image of a mock $^{nat}Sr(NO_3)_2$ target sample prepared under conditions identical to actual targets. The analysis revealed that the typical diameter of individual dots was approximately 100 μm, corresponding to a surface coverage of approximately 7%. Therefore, the dot thickness was approximately 14 times the average thickness derived from the $^{85}Sr$ radioactivity measurement (calculated as 1/0.07). Consequently, the simulation was executed as follows: for each layer of the dot target, the energy loss was calculated assuming a thickness 14 times the nominal value with a 7% probability, whereas the calculation was skipped with a 93% probability. In the present experiment, the estimated energy loss based on this local thickness remained below 0.1 MeV, even for the target with the lowest incident energy. Furthermore, the energy spread attributed to the target inhomogeneity, estimated from the dispersion of the Monte Carlo simulations, was less than 0.03 MeV. These contributions accounted for approximately 10% of the total uncertainty in the projectile energy for each target.

After irradiation, γ-rays emitted from the target foils were measured using four germanium detectors. The energy and detection efficiency of the detectors were calibrated using a standard γ-ray point source (Type M GF-ML-7601, Eckert & Ziegler Isotope Products). The energy resolutions of the detectors ranged in 1.8–1.9 keV of FWHM at 1332 keV. To ensure the quantitative collection of recoiled reaction products, each $^{nat}Sr(NO_3)_2$ target was measured together with its corresponding $^{27}Al$ cover and backing foils. Measurements were performed at more than ten different cooling times for each sample to follow the decay curves and to verify the half-lives of the produced radionuclides.

The reaction cross sections were derived from the induced radioactivities at the end of bombardment. For each characteristic γ-ray peak, the radioactivity at the end of bombardment was determined by fitting a decay curve to multiple measurement points using a weighted least-squares method. These radioactivities had statistical uncertainties of typically less than 10%. For the multiple intense γ-ray emitting nuclides, $^{86g,87g,88}Y$, the final radioactivity was determined as the weighted average of the values obtained from the individual γ-ray peaks. In this averaging process, the uncertainty of radioactivity obtained from each peak was given as the root of square sum of the uncertainties of decay-curve fitting, γ-ray emission probability (<3.1%), and detection efficiency at corresponding γ-ray energy (5%).

The total uncertainty of the cross sections was evaluated as the root of square sum of the uncertainties in the induced radioactivity and the target thickness. The uncertainty in the target thickness was derived from the root of square sum of the uncertainties of counting statistics (1.7–3.3%), detector efficiency (5%) for the 514-keV γ-ray of $^{85}Sr$, and spatial inhomogeneity evaluated by autoradiography (3%) as described in Appendix B.

Nuclear decay data essential for the analysis were retrieved from NuDat 3.0 [37] and the IAEA Live Chart of Nuclides [38], except the recent references for $^{86}Y$ [39]. The theoretical cross-section data, Q-value, and threshold energy were obtained from TENDL-2023 [32]. The decay data used in this work are summarized in Table 2. To validate the beam parameters and target thicknesses, the cross sections of the monitor reaction $^{nat}Cu(d,x)^{65}Zn$ were determined. Figure 3 compares our experimental results with the IAEA recommended values [40]. The measured cross sections show good agreement with the recommended data over the entire energy range investigated. Although the measured values were slightly higher than the recommended values in the middle energy region (10–15 MeV), they remained consistent within experimental uncertainties (approx. 1σ). Consequently, no adjustments were made to the initial beam parameters or target thicknesses.

Table 1. Summary of experimental parameters: target specifications, measured beam parameters, and measurement conditions.

| | |
|---|---|
| **Target** | |
| Configuration | $X_1$-$Y_1$-$X_2$-$Y_2$-…-$X_{15}$-$Y_{15}$-$X_{16}$ |
| | $X_i$ = $^{27}$Al-$^{nat}$Cu-$^{27}$Al |
| | $Y_i$ = $^{27}$Al(cover)-$^{nat}Sr(NO_3)_2$-$^{27}$Al(backing) |
| $^{nat}Sr(NO_3)_2$ | 72±5 μg/cm$^2$ ($i$ = 1–8), 92±6 μg/cm$^2$ ($i$ = 9–15) |
| $^{nat}$Cu | 21.7±0.4 mg/cm$^2$ |
| $^{27}$Al | 2.75±0.06 mg/cm$^2$, 2.61±0.05 mg/cm$^2$ |
| **Beam** | |
| Energy | 24.5±0.1 MeV |
| Average intensity | 99.3 enA |
| Irradiation period | 30 min |
| **Measurement** (cooling time, measurement time, distance, dead time, and number of analyzed spectra for each product) | |
| $^{86m}$Y | 37–203 min, 5–10 min, 2–20 cm, <10%, 2 |
| $^{86g}$Y | 0.15–5 d, 3 h, 10 cm, 0–7%, 3–6 |
| $^{87m}$Y | 0.15–5 d, 3 h, 10 cm, 0–7%, 4–7 |
| $^{87g}$Y | 5–13 d, 6–20 h, 10 cm, ~0%, 1–3 |
| $^{88}$Y | 5–24 d, 6 h–3 d, 10 cm, ~0%, 3–4 |

Table 2. Reaction and decay data of product nuclides. The contributing reactions with the smallest Q-values were listed among the possible reactions to form the given nuclide.

| Nuclide | Half-life | Decay mode (%) | $E_\gamma$ (keV) | $I_\gamma$(%) | Contributing reaction | Q-value (MeV) | Threshold energy (MeV) |
|---|---|---|---|---|---|---|---|
| $^{86m}$Y | 47.4(4) min | IT 99.31(4) | 208.1 | 93.8 | $^{84}$Sr($d$,$\gamma$) | 11.6 | 0 |
| | | $\beta^+$ 0.44(3) | | | $^{86}$Sr($d$,2$n$) | -8.5 | 8.7 |
| | | EC 0.25(3) | | | $^{87}$Sr($d$,3$n$) | -16.9 | 17.3 |
| $^{86g}$Y | 14.74(2) h | EC 72.8(20)[a] | 443.1 | 16.9(5) | $^{84}$Sr($d$,$\gamma$) | 11.8 | 0 |
| | | $\beta^+$ 27.2(20)[a] | 627.7 | 32.6(10) | $^{86}$Sr($d$,2$n$) | -8.2 | 8.4 |
| | | | 777.4 | 22.4(6) | $^{87}$Sr($d$,3$n$) | -16.7 | 17.1 |
| | | | 1076.6 | 82.5 | $^{86m}$Y decay | | |
| $^{87m}Y$ | 13.37(3) h | IT 98.43(11) | 380.8 | 78.05 | $^{86}Sr(d,n)$ | 3.2 | 0 |
| | | EC 0.82(5) | | | $^{87}Sr(d,2n)$ | -5.2 | 5.4 |
| | | $\beta^+$ 0.75(5) | | | $^{88}Sr(d,3n)$ | -16.4 | 16.7 |
| $^{87g}Y$ | 79.8(3) h | EC 99.820(20) | 388.5$^b$ | 82.2$^b$ | $^{86}Sr(d,n)$ | 3.6 | 0 |
| | | $\beta^+$ 0.180(20) | 484.8 | 89.8(9) | $^{87}Sr(d,2n)$ | -4.9 | 5.0 |
| | | | | | $^{88}Sr(d,3n)$ | -16.0 | 16.3 |
| | | | | | $^{87m}Y$ decay | | |
| $^{88}Y$ | 106.6(21) d | EC 99.79 | 898.0 | 93.7(3) | $^{86}Sr(d,\gamma)$ | 12.9 | 0 |
| | | $\beta^+$ 0.21 | 1836.1 | 99.2(3) | $^{87}Sr(d,n)$ | 4.5 | 0 |
| | | | | | $^{88}Sr(d,2n)$ | -6.6 | 6.8 |
| $^{65}Zn$ | 243.93(9) d | EC 98.579(7) | 1115.5 | 50.04(10) | $^{63}Cu(d,\gamma)$ | 13.5 | 0 |
| | | $\beta^+$ 1.421(7) | | | $^{65}Cu(d,2n)$ | -4.4 | 4.5 |

*a*: Taken from Ref. [39].

*b*: Emitted from the daughter nuclide, $^{87m}Sr$ ($T_{1/2}$ = 2.82 h).

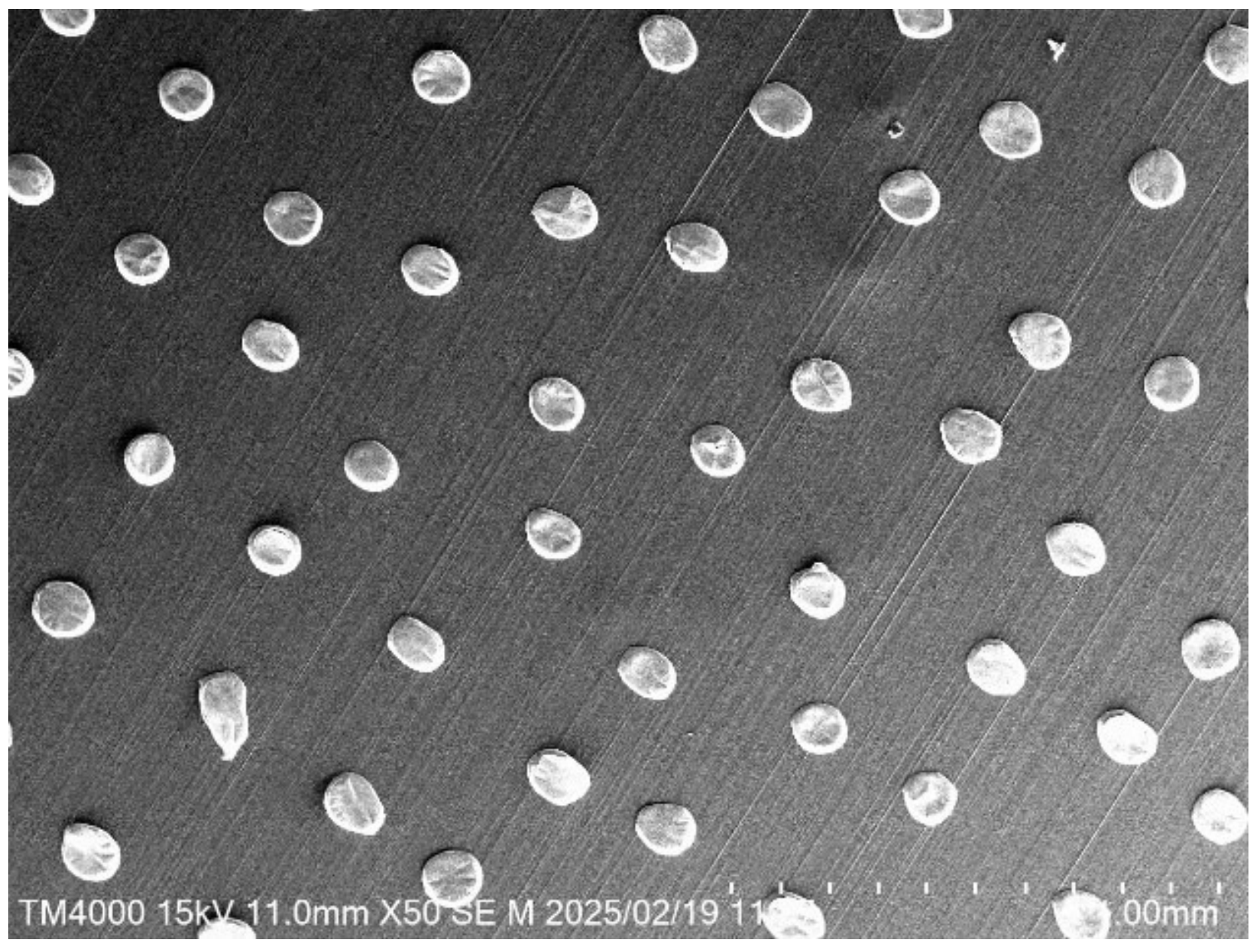

Fig. 2. The SEM image of a mock strontium sample prepared under conditions identical to the actual target. Scale bar: 1 mm across 10 tick marks.

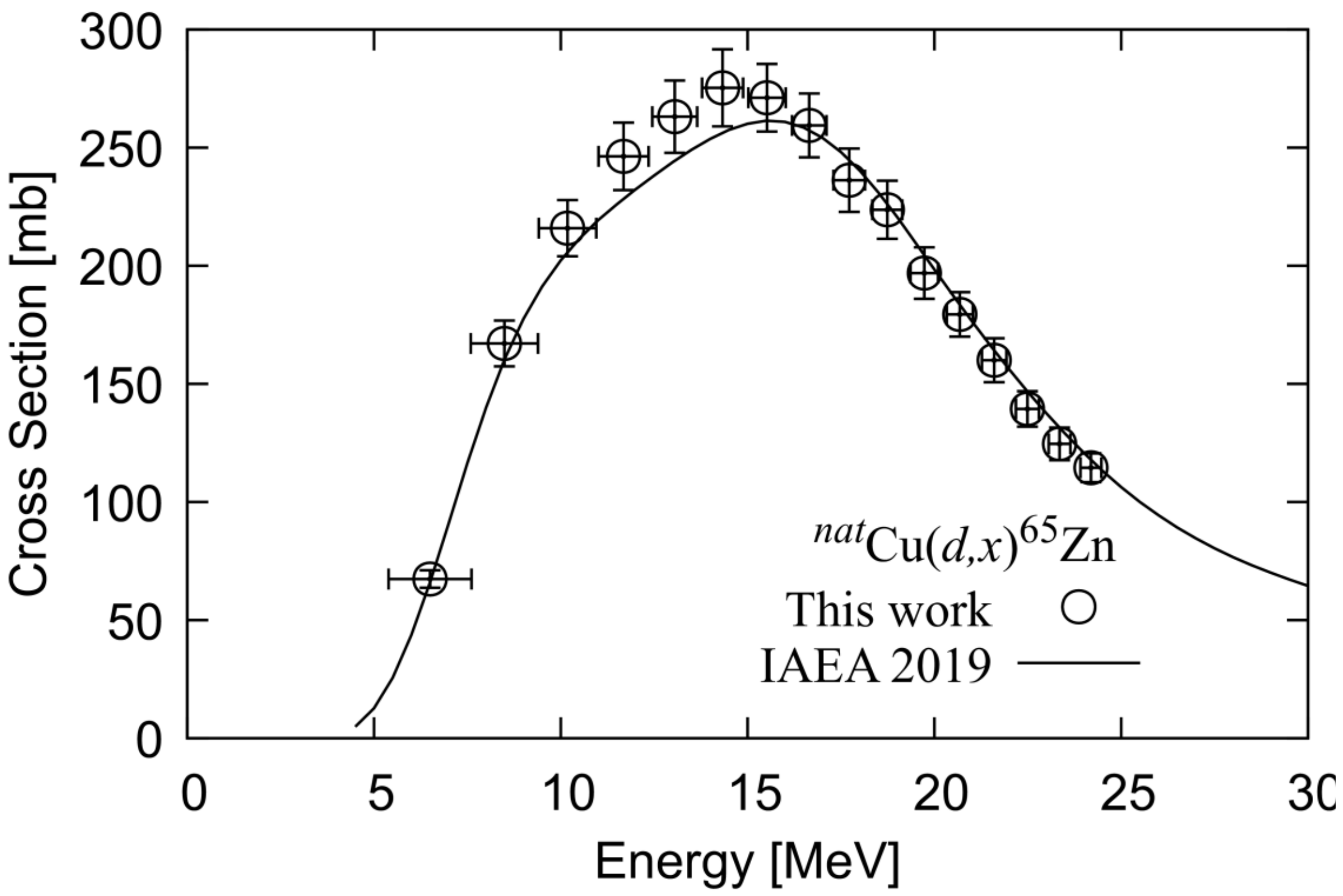

Fig. 3 Cross sections of the ${}^{nat}$Cu($d$,$x$)${}^{65}$Zn monitor reaction with the IAEA recommended values [40].

Result and discussion

The production cross sections for $^{86m,g}Y$, $^{87m,g}Y$, and $^{88}Y$ were experimentally determined in the energy range up to 24 MeV. The obtained numerical data are listed in Table 3. In Figs. 4–8, the present results are compared with the literature data [27] and theoretical predictions from the TENDL-2023 library [32]. The theoretical values for the $^{nat}Sr$ target were calculated by taking the abundance-weighted sum of the cross sections for the constituent isotopes: $^{84}Sr$ (0.56%), $^{86}Sr$ (9.86%), $^{87}Sr$ (7.0%), and $^{88}Sr$ (82.6%) [37].

Table 3. Activation cross sections of deuteron-induced reactions on $^{nat}Sr$.

| Energy (MeV) | $^{86m}Y$ (mb) | $^{86g}Y$ (mb) | $^{87m}Y$ (mb) | $^{87g}Y$ (mb) | $^{88}Y$ (mb) |
|---|---|---|---|---|---|
| 23.8±0.1 | 45±5 | 103±11 | 420±33 | 650±50 | 488±35 |
| 22.9±0.1 | 44±4 | 101±10 | 357±30 | 560±40 | 520±40 |
| 22.1±0.2 | 40±4 | 97±12 | 280±22 | 439±32 | 610±40 |
| 21.1±0.2 | 34±4 | 98±11 | 239±20 | 367±28 | 690±50 |
| 20.2±0.2 | 36±4 | 91±9 | 169±14 | 272±20 | 770±60 |
| 19.2±0.2 | 34.3±3.4 | 94±8 | 125±10 | 195±14 | 870±60 |
| 18.2±0.3 | 32.5±3.2 | 87±10 | 80±7 | 123±10 | 910±70 |
| 17.2±0.3 | 25.9±3.3 | 86±8 | 67±6 | 99±8 | 860±60 |
| 16.1±0.3 | 23.9±3.0 | 83±8 | 66±6 | 95±7 | 850±60 |
| 14.9±0.4 | 21.6±2.9 | 74±7 | 66±6 | 100±8 | 810±60 |
| 13.7±0.4 | 16.9±1.4 | 59±5 | 50±4 | 79±6 | 630±50 |
| 12.4±0.5 | 10.4±1.0 | 40.1±3.2 | 47±4 | 71±5 | 530±40 |
| 11.0±0.6 | 6.4±0.5 | 25.8±2.0 | 48±4 | 77±6 | 450±34 |
| 9.4±0.7 | 0.85±0.10 | 5.1±0.6 | 46±4 | 80±6 | 295±22 |
| 7.6±0.8 | — | — | 25.1±2.2 | 50±4 | 62±6 |

$^{86m}$Y production

The production cross sections of the metastable state $^{86m}$Y ($T_{1/2}$ = 47.4 min) were determined by analyzing the prominent γ-ray peak at 208.1 keV ($I_{\gamma}$ = 93.8%). Due to the short half-life, the analysis relied on the spectra obtained during the first measurement period (shortest cooling time). Figure 4 presents the obtained excitation function in comparison with the theoretical predictions from TENDL-2023 [32]. To the best of our knowledge, this study provides the first experimental dataset for the reaction. The experimental values are systematically lower than the theoretical predictions by approximately 20%.

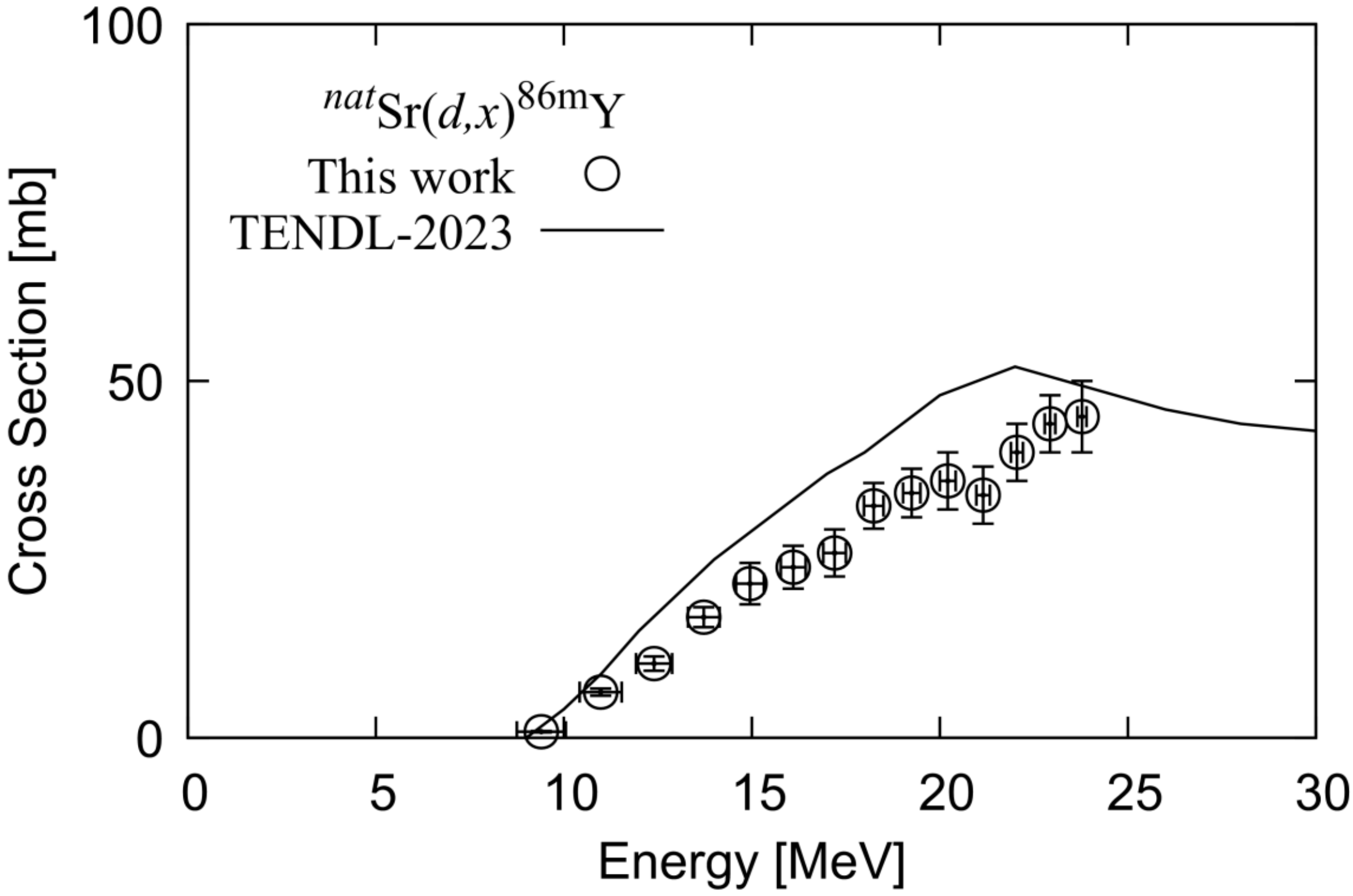

Fig. 4. Cross sections of the $^{nat}$Sr($d$,$x$)$^{86m}$Y reaction with the TENDL-2023 values [32].

$^{86g}$Y production

The cumulative production cross sections of $^{86g}$Y ($T_{1/2}$ = 14.74 h) were determined by individually analyzing the intense γ lines at 443.1, 627.7, 777.4, and 1076.6 keV ($I_{\gamma}$ = 16.9, 32.6, 22.4, and 82.5%, respectively). Data acquired after a cooling time of more than 3 hours were adopted for the analysis. By this time, the precursor $^{86m}$Y had decayed to the ground state sufficiently, reducing its contribution to less than 1% of the measured $^{86g}$Y radioactivity: consequently, the determined cross sections represent the cumulative formation ($^{86m+g}$Y). The results derived from each γ-ray line were consistent. The cross sections were finally determined as the weighted average of the radioactivities derived from these γ lines.

Figure 5 shows our results in comparison with the previous study [27] and the theoretical calculations [32]. The data reported by Tárkányi *et al.* [27] are consistent with our results above 13.5 MeV, however, the data at 11.7 MeV is lower than our data. The theoretical prediction from the TENDL-2023 library [32] agrees with experimental values within error above 21 MeV, while it systematically overestimates the experimental values by approximately 20% below 21 MeV.

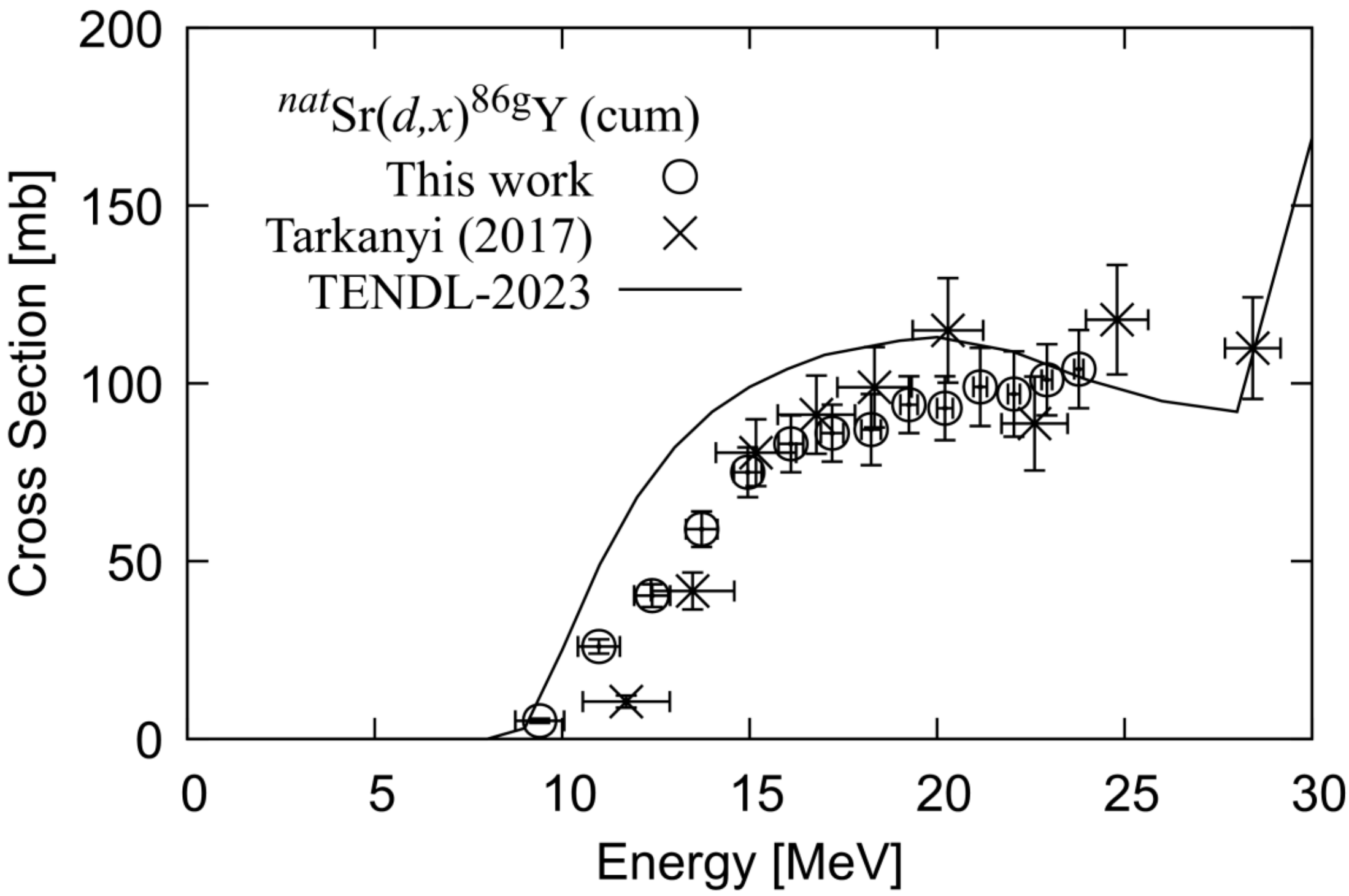

Fig. 5. Cumulative cross sections of the $^{nat}$Sr($d,x$)$^{86g}$Y reaction with the experimental literature data [27] and the TENDL-2023 values [32].

$^{87m}$Y production

The independent production cross sections for $^{87m}$Y ($T_{1/2}$ = 13.37 h) were determined using the prominent γ line at 380.8 keV ($I_{\gamma}$ = 78.05%). Results derived from multiple measurements showed good consistency. The final values were adopted as the weighted mean of the data obtained from spectra measured within three days after the end of bombardment (EOB).

Figure 6 displays our results in comparison with the previous study [27] and the theoretical calculations [32]. The cross sections reported by Tárkányi *et al.* [27] are higher than ours above 20.3 MeV, but lower below 9.8 MeV. The theoretical prediction given by TENDL-2023[32] overestimates our results over the entire energy range, though it agrees with the experimental data by Tárkányi *et al.* [27] above 15 MeV.

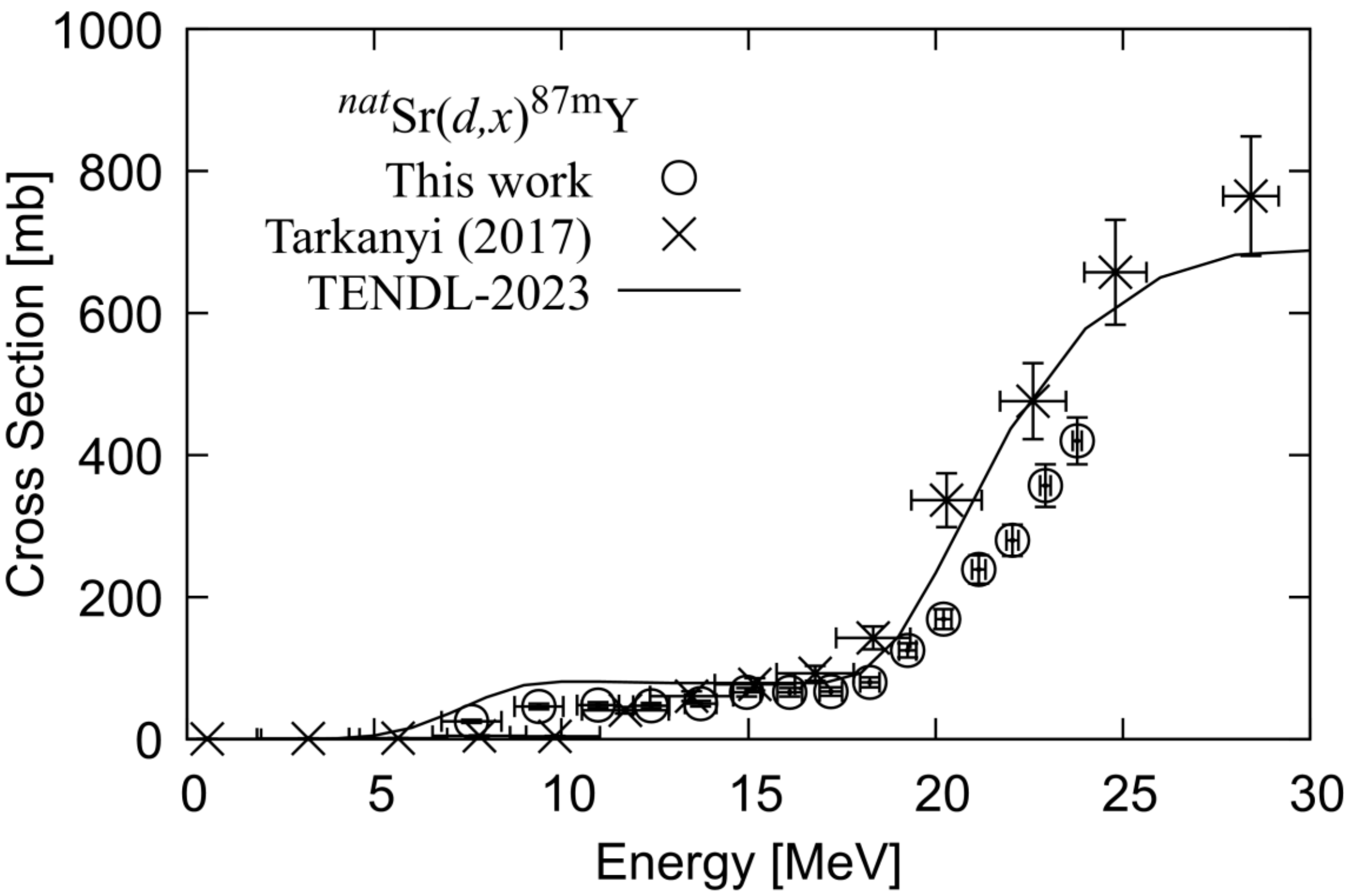

Fig. 6. Cross sections of the $^{nat}$Sr($d$,$x$)$^{87m}$Y reaction with the experimental literature data [27] and the TENDL-2023 values [32].

$^{87g}$Y production

The cumulative production cross sections of $^{87g}$Y ($T_{1/2}$ = 79.8 h) were determined utilizing the intense γ lines at 388.5 and 484.8 keV ($I_{\gamma}$ = 82.2 and 89.8%, respectively). The data acquired after a cooling time of 5–10 days were adopted for the analysis. By this time, the precursor $^{87m}$Y ($T_{1/2}$ = 13.37 h, IT = 98.43%) decayed sufficiently to the ground state $^{87g}$Y, and its contribution was estimated to be less than 1% of the measured radioactivity of $^{87g}$Y: thus, the determined cross sections are cumulative. Regarding the 388.5-keV line, which is emitted from the daughter nuclide $^{87m}$Sr ($T_{1/2}$ = 2.8 h), the contribution from directly produced $^{87m}$Sr was negligible after the long cooling period. Since transient equilibrium was established between $^{87g}$Y and $^{87m}$Sr, the radioactivity of $^{87g}$Y was directly estimated from that of $^{87m}$Sr. The results derived from the two γ lines were consistent. Finally, the representative cross sections were obtained as the weighted average of these results.

Figure 7 shows the present results in comparison with the previous study [27] and the theoretical calculations [32]. The data reported by Tárkányi *et al.* [27] are higher than our results above 15 MeV but lower below 9.8 MeV. The theoretical prediction from the TENDL-2023 library[32] is generally consistent with our experimental values.

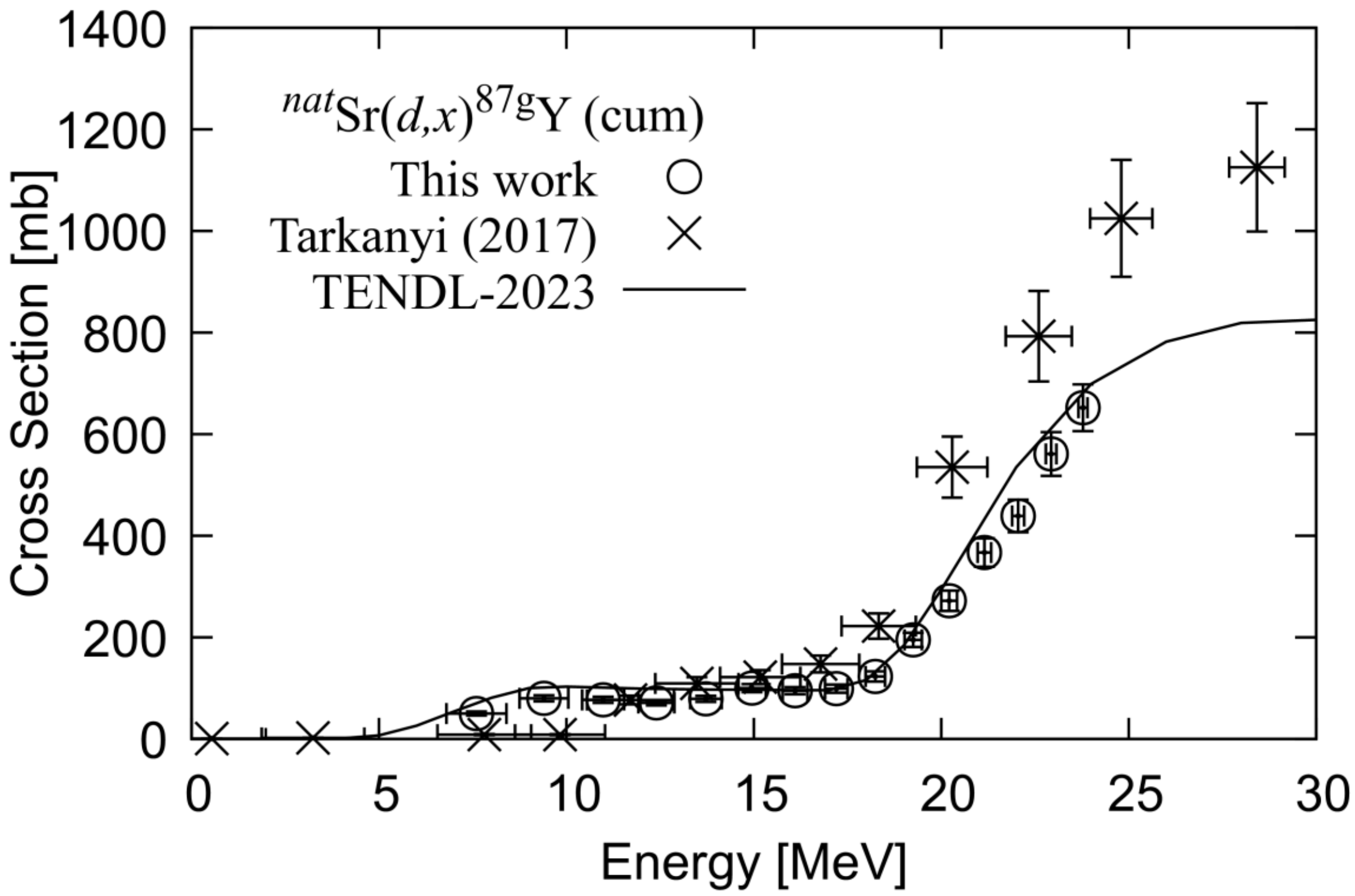

Fig. 7. Cumulative cross sections of the $^{nat}$Sr($d,x$)$^{87g}$Y reaction with the experimental literature data [27] and the TENDL-2023 values [32].

$^{88}$Y production

The production cross sections for $^{88}$Y ($T_{1/2}$ = 106.6 d) were determined by analyzing the two prominent γ lines at 898.0 keV ($I_\gamma$ = 93.7%) and 1836.1 keV ($I_\gamma$ = 99.2%). Data obtained after a cooling period of 13 days were selected for the analysis. The results derived from these two independent lines showed consistency. Consequently, the weighted average of the values was adopted as the representative cross section.

Figure 8 presents our results alongside the previous experimental data [27] and the theoretical calculations [32]. The data reported by Tárkányi *et al.* [27] show good agreement with the present results in the energy region above 13 MeV. However, they are systematically lower below 13 MeV. The theoretical predictions from the TENDL-2023 library [32] overestimate the experimental values throughout the measured energy range.

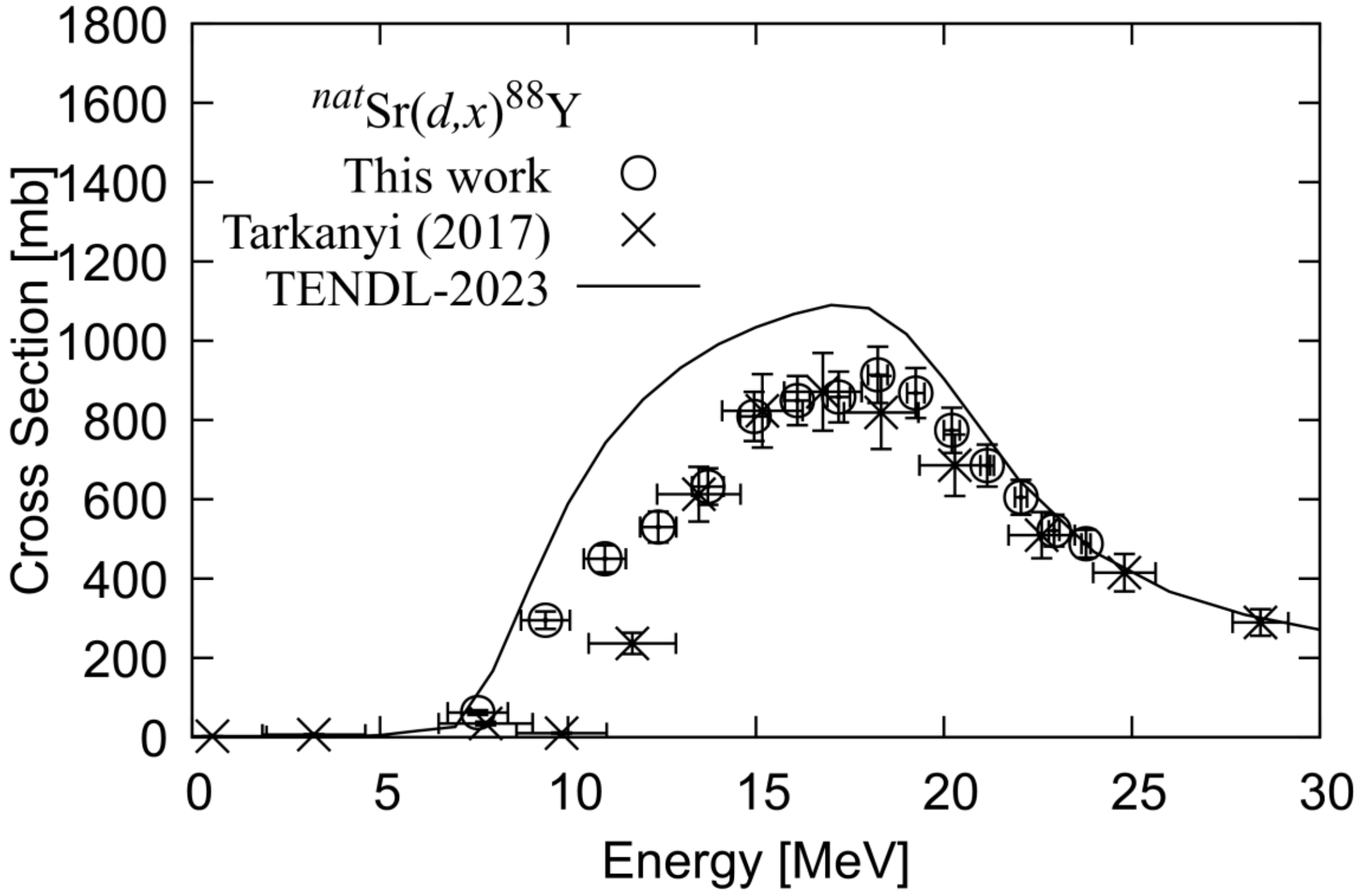

Fig. 8. Cross sections of the $^{nat}$Sr($d,x$)$^{88}$Y reaction with the experimental literature data [27] and the TENDL-2023 values [32].

## Conclusion

We determined the production cross sections for yttrium isotopes via the deuteron-induced reactions on $^{nat}$Sr. The inkjet printing technique was successfully utilized to prepare practical thin-layer targets for cross section measurements, demonstrating its applicability for target elements whose metal foil is unavailable. This study notably provides the first experimental data for $^{86m}$Y. Comparisons with TENDL-2023 revealed a general overestimation by the theoretical code, suggesting a need for parameter adjustments. Regarding the production perspectives, the investigated energy range corresponds to the optimal window for the production of $^{88}$Y. For $^{86g}$Y and $^{87g}$Y, although higher energy ranges are generally preferred for effective production, the present low-energy data serve as an important benchmark for validating nuclear reaction codes and assessing impurity levels. These results collectively improve the reliability of nuclear data for applications.

## Acknowledgement

This work was carried out at RI Beam Factory operated by RIKEN Nishina Center and CNS, University of Tokyo, Japan. This research was partially performed by the commissioned research fund provided by F-REI (JPFR25040201) and JSPS KAKENHI Grant Number 22H04961.

## CRediT

AN: Project administration, Formal analysis, Investigation, Writing - original draft. AM: Investigation, Writing – original draft, Supervision. YS, YK, TT, SM: Investigation, Writing - review & editing. HH: Investigation, Resources, Funding acquisition, Writing – original draft, Supervision.

## Appendix

### Appendix A. Preparation of $^{nat}Sr(NO_3)_2$ solution

Two independent batches of $^{nat}Sr(NO_3)_2$ solution were prepared in 0.01 M nitric acid as shown in Fig. A.1.

For batch #1, 192.64 mg of vacuum-dried $^{nat}SrCO_3$ powder (purity: 99.99%, Wako Pure Chemical Industries, Ltd., Japan) was weighed with precision and placed in a glass beaker. To quantify the amount of $^{nat}$Sr deposited, 20 μL of a $^{85}$Sr solution (87.5 kBq at the date of target fabrication, 0.01 M nitrate acid) was added to the beaker as a yield tracer. The no-carrier-added $^{85}$Sr was produced via the $^{nat}$Rb($d,x$)$^{85}$Sr reaction and chemically separated from the $^{nat}$RbCl target [41]. Due to the low solubility of $^{nat}Sr(NO_3)_2$ in nitric acid, the mixture was first dissolved using hydrochloric acid (FUJIFILM Wako Pure Chemical Corporation, 081-03475; for Analysis of Poisonous Metals). To prevent reagent loss due to vigorous dissolutions, the acid was initially diluted and added dropwise in several stages, with each addition limited to a few hundred microliters, and the acid concentration was gradually increased. After complete dissolution of the $SrCO_3$ powder, the solution was evaporated to dryness and redissolved in Milli-Q water (18 MΩ cm, Milli-Q Advantage A10; Merck KGaA, Germany) and 1 M nitric acid (FUJIFILM Wako Pure Chemical Corporation, 148-03515; for Volumetric Analysis). This evaporation and redissolution step was repeated three times to fully convert the chemical form chloride to nitrate. The resulting residue was then dissolved in 800 µL of 0.01 M nitric acid (KISHIDA CHEMICAL CO., LTD., 900-02045) and transferred from the glass beaker to a 10-mL volumetric flask. The glass beaker was rinsed five times with totally 800-µL aliquots of 0.01 M nitric acid, and the washings were added to the same volumetric flask. Finally, the solution was diluted to a total volume of 10 mL with 0.01 M nitric acid. The batch #1 contained 11.43 mg/mL of $^{nat}Sr^{2+}$ and 8.75 kBq/mL of $^{85}Sr$ (specific radioactivity: 765 Bq/mg).

For batch #2, a 10-mL non-spiked $^{nat}Sr(NO_3)_2$ solution was prepared in advance using a similar procedure. A 755-µL aliquot of this solution was transferred via a micropipette to a glass vial and mixed with 0.25 mg of the spike solution ($^{85}Sr$ 3.97 kBq, 0.01 M nitric acid).

The batch #2 contained 11.48 mg/mL of $Sr^{2+}$ and 5.16 kBq/mL of $^{85}Sr$ (specific radioactivity: 449 Bq/mg).

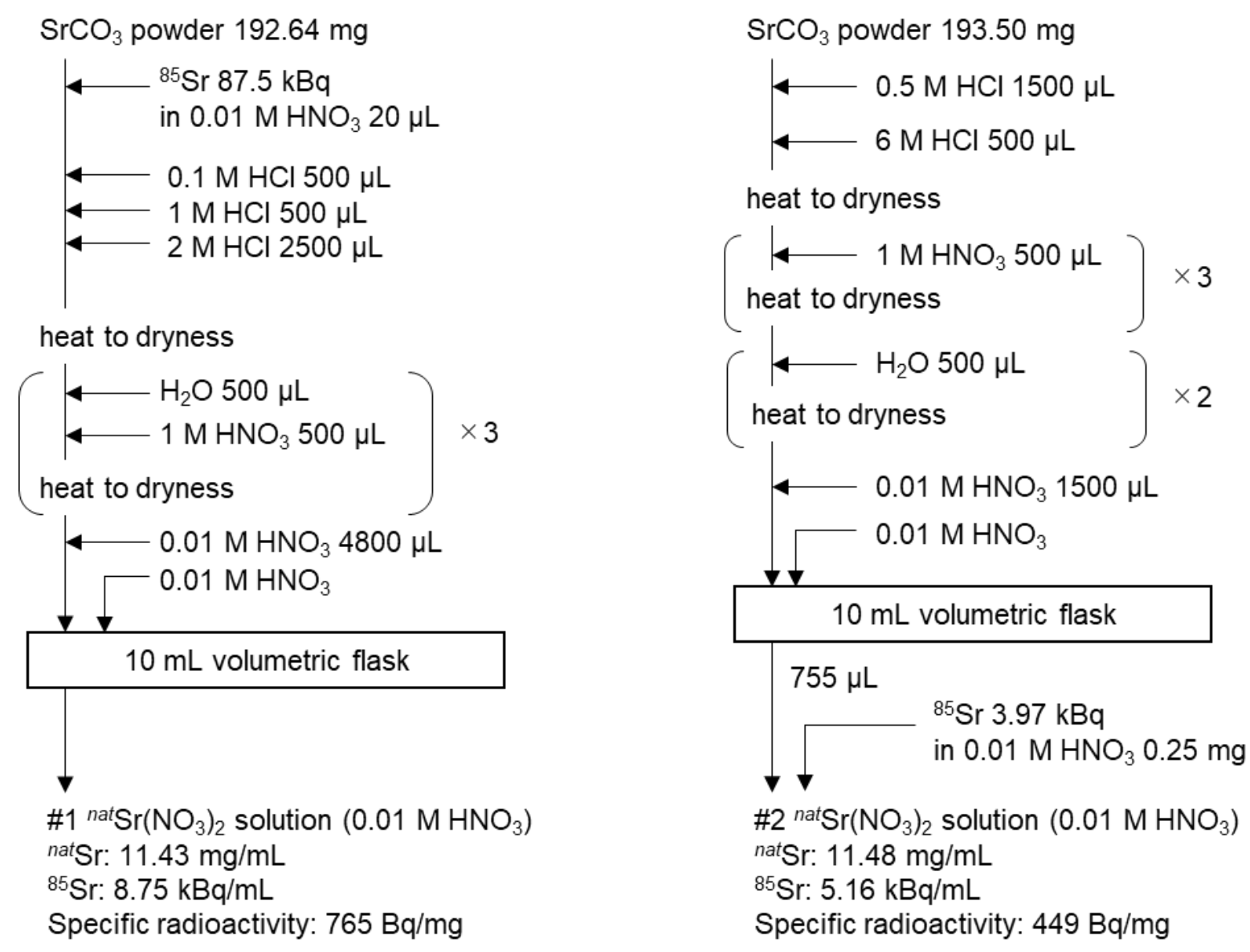

Fig. A.1. Chemical procedure for preparing two batches of the $^{nat}Sr(NO_3)_2$ solution.

## Appendix B. Evaluation of uniformity of $^{nat}Sr(NO_3)_2$ target thickness

The targets fabricated using the inkjet printer consisted of an array of localized thick dots distributed over a wide area (see Fig. 2 in the main text). If the beam center aligns with a target dot, the interaction probability between the projectile and target nuclei can increase, as the actual beam intensity is also spatially non-uniform. This variation in the interaction probability is analogous to the effects caused by thickness fluctuations in conventional uniform-thickness targets. In this study, the impact of this spatial inhomogeneity was quantitatively evaluated by analyzing autoradiographic images under assumptions of potential beam profiles. Finally, this effect was treated as a random uncertainty component of the target thickness.

The spatial inhomogeneity of the target thickness across the printed area was evaluated through the following steps: (1) calculating the mean PSL intensity, (2) modeling the incident beam profile, (3) evaluating the spatial inhomogeneity of the PSL intensity, and (4) subsequently converting these results into the spatial inhomogeneity of target thickness.

(1) Mean PSL intensity

In autoradiography, photostimulated luminescence (PSL) intensity was acquired as an 8-bit (256-level) integer for each pixel, in a spatial resolution of 50 μm, and the whole data was saved as a grayscale bitmap image. For the $i$-th sample, the analysis region, $A^{(i)}$, was manually selected using ImageJ software [42] as presented in Fig. B.1. Let $L_p^{(i)}$ denote the PSL intensity of a single pixel $p$ within $A^{(i)}$; the mean PSL intensity per pixel, $L_{\text{mean}}^{(i)}$, within $A^{(i)}$ is calculated using Eq. (B.1):

$$L_{\text{mean}}^{(i)} = \frac{1}{|A^{(i)}|} \sum_{p \in A^{(i)}} L_p^{(i)} \tag{B.1}$$

where $|A^{(i)}|$ represents the number of pixels within the region $A^{(i)}$. This $L_{\text{mean}}^{(i)}$ corresponds to the conventional target thickness.

(2) Incident beam profile

The incident beam profile was modeled as a Gaussian distribution shaped by the collimator aperture. The width of the Gaussian distribution was determined from the beam current ratio between the target and the $\phi$ 3-mm collimator. The analysis was initially conducted using the beam current ratio of 0.5:0.5, which was monitored during the beam tuning immediately before target irradiation. Although fluctuation in the beam current on the target was within 3%, the analysis was conservatively repeated across a broader range from 0.3:0.7 to 0.7:0.3. For each assumed beam current ratio, the Gaussian beam profile was approximated as a step-wise distribution as the weighted sum of three uniform circular components with diameters of 1, 2, and 3 mm as shown in Fig. B.2. These specific components were subsequently utilized in the following analytical procedure.

(3) Spatial inhomogeneity of PSL intensity

Because the local target thickness is proportional to the PSL intensity, the spatial inhomogeneity of each $^{nat}$Sr target thickness was primarily evaluated in terms of the PSL intensity using a Monte Carlo analysis; for each of the 1-, 2-, and 3-mm diameter beam components, 1,000 fixed-size circular regions of interest (ROIs) were randomly distributed within the $\phi$ 6-mm area from target center to cover potentially irradiated region. A representative set of 50 sampled ROIs and the frequency distribution of the analyzed thickness from the 1,000 calculations for the $\phi$ 3-mm diameter ROI on the first $^{nat}$Sr target ($i = 1$) are shown in Figs. B.3 and B.4, respectively. Within each ROI, mean PSL intensity per pixel was calculated. Subsequently, its mean and standard deviation among 1,000 iterations of Monte Carlo analysis, $L_{\mathrm{MC}}^{(i)}$ and $\Delta L_{\mathrm{MC}}^{(i)}$, respectively, were derived. These values for Gaussian-shaped beam were derived by the weighted average of those for $\phi$ 1, $\phi$ 2, and $\phi$ 3 mm, where the weights of them (2.1, 13.4, and 84.6%, respectively) were mathematically derived from their volume fractions under assumption illustrated in Fig. B.2. Let $\Delta L_{\mathrm{sys}}^{(i)}$ denote the systematic error of $L_{\mathrm{MC}}^{(i)}$ compared from $L_{\mathrm{mean}}^{(i)}$ and $\Delta L_{\mathrm{stat}}^{(i)}$ denote its statistic error; they are expressed in Eq. (B.2) and (B.3), respectively.

$$\Delta L_{\mathrm{sys}}^{(i)} = L_{\mathrm{MC}}^{(i)} - L_{\mathrm{mean}}^{(i)} \qquad \text{(B.2)}$$

$$\Delta L_{\mathrm{stat}}^{(i)} = \Delta L_{\mathrm{MC}}^{(i)} \qquad \text{(B.3)}$$

The representative value for 15 $^{nat}$Sr targets was derived as the root of sum square for those results.

(4) Spatial inhomogeneity of target thickness

Consequently, the relationship between PSL intensity and radioactivity of the $^{nat}$Sr target was estimated to determine the uncertainty of the thickness of each target. The radioactivity of the *i*-th target was measured using a germanium detector. The radioactivity of the *i*-th $^{nat}$Sr target, $D^{(i)}$, is assumed to be proportional to the background-corrected PSL intensity as expressed in Eq. (B.4).

$$D^{(i)} = c\left(L_{\mathrm{mean}}^{(i)} - L_{\mathrm{BG}}\right) \qquad \text{(B.4)}$$

Here *c* is a conversion coefficient from PSL intensity to radioactivity and $L_{\mathrm{BG}}$ is the PSL intensity originated from background, which were estimated by extrapolating the linear relationship between radioactivity and mean PSL intensity among 15 $^{nat}$Sr targets, as illustrated in Fig. B.5. Systematic and statistic uncertainty of $D^{(i)}$ were derived from $\Delta L_{\mathrm{sys}}^{(i)}$ and $\Delta L_{\mathrm{stat}}^{(i)}$ by error propagation; $\Delta D_{\mathrm{sys}}^{(i)} = c\Delta L_{\mathrm{sys}}^{(i)}$ and $\Delta D_{\mathrm{stat}}^{(i)} = c\Delta L_{\mathrm{stat}}^{(i)}$. Because $^{nat}$Sr target thickness is proportional to $^{85}$Sr radioactivity, relative uncertainty of *D* presents the relative uncertainty of target thickness. The numerical results of $\Delta D^{(i)}/D^{(i)}$ are listed in Table B.1 with total relative uncertainty, δ, defined as root sum square of its relative systematic uncertainty and relative statistic uncertainty. Finally, δ consistently remained below 3% under all examined conditions; consequently, a fixed value of 3% was adopted as the uncertainty attributed to the spatial inhomogeneity of the Sr target thickness.

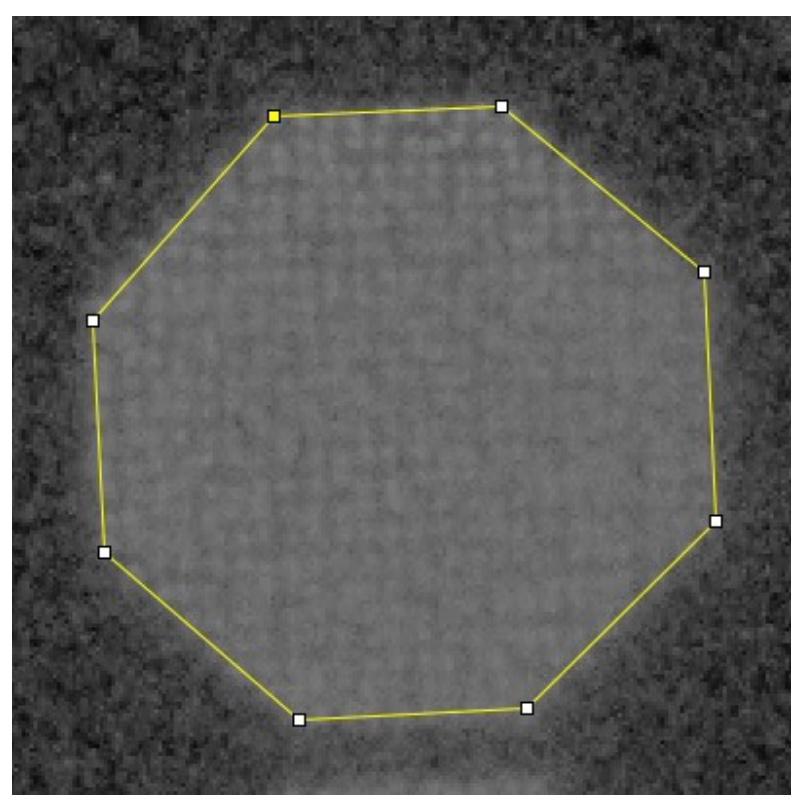

Fig. B.1. A typical example of analysis region $A^{(i=1)}$ manually set on the $^{nat}$Sr target ($i = 1$).

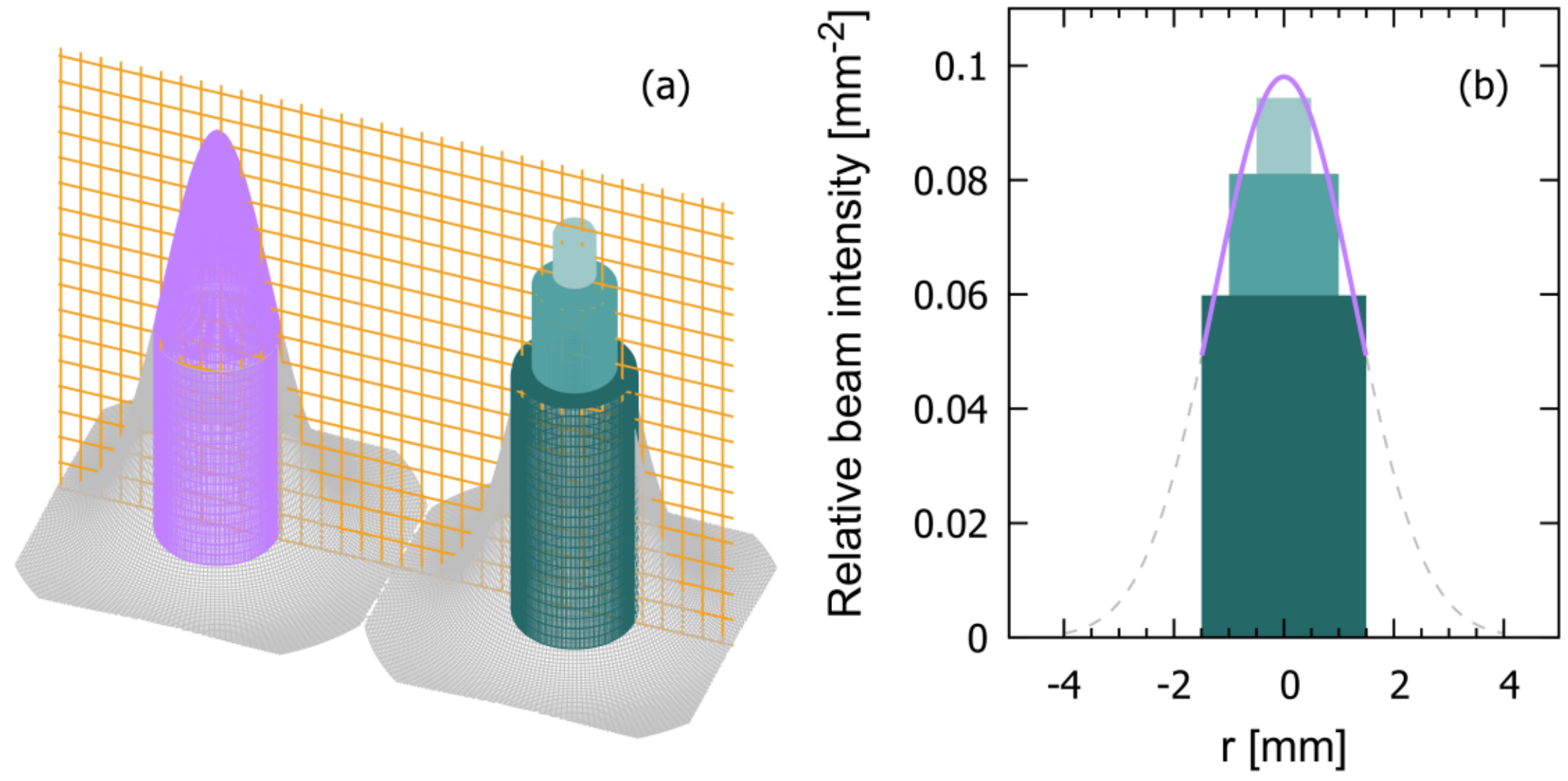

Fig. B.2. Illustrations of the estimated Gaussian beam profile and its approximation using uniform $\phi$ 1-, 2-, and 3-mm circular components. (a) a 3D view of the individual profiles and (b) an overlaid 2D cross-sectional view. The gray-shaded region outside of the $\phi$ 3-mm ($|r| >$1.5 mm) from the target center represents the portion intercepted by the collimator.

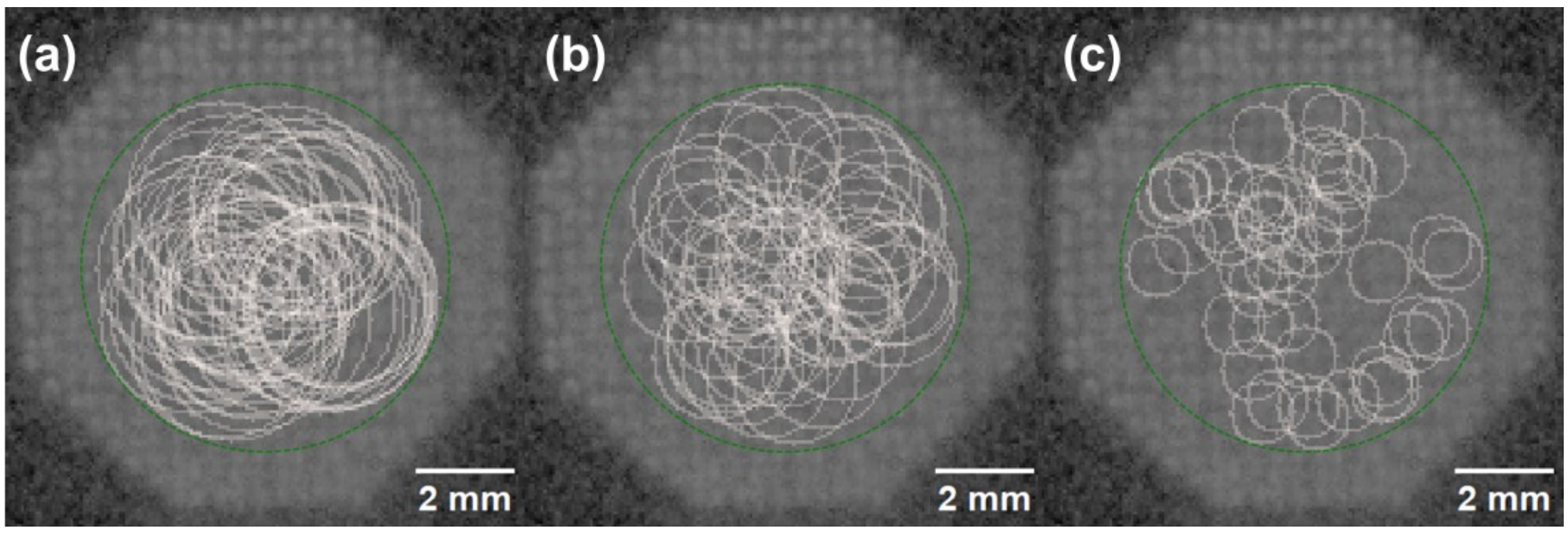

Fig. B.3. Visual illustrations of a representative set of 50 sampled ROIs generated during the 1,000 random-spot analysis for the first Sr target ($i = 1$). Panels (a-c) correspond to $\phi$ 3-, $\phi$ 2-, and $\phi$ 1-mm ROIs, respectively.

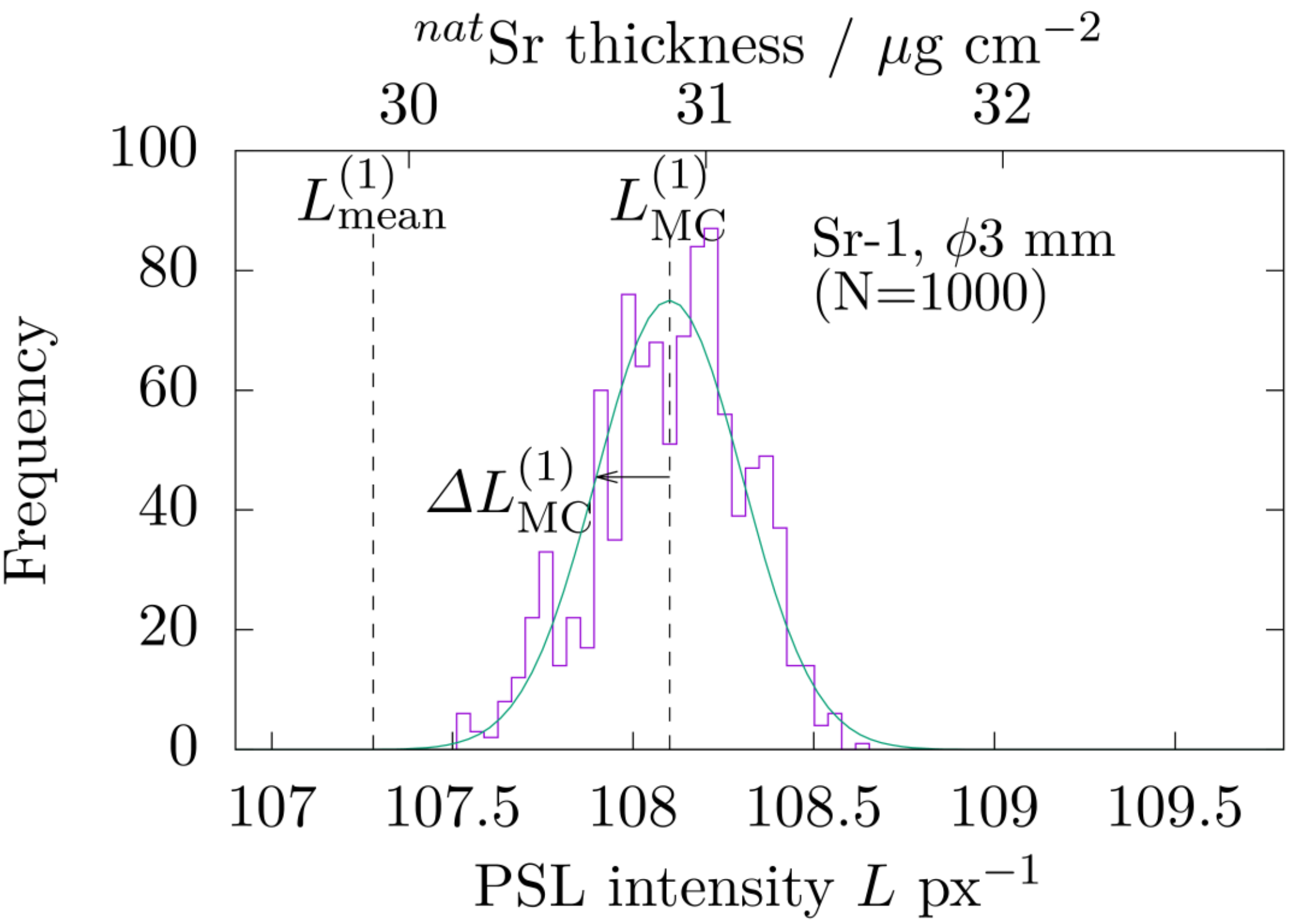

Fig. B.4. Distribution of the mean PSL intensity per pixel ($L$) within a $\phi$ 3-mm ROI ($m = 3$) for the first Sr target ($i$=1), obtained from 1,000 random iterations. The mean and standard deviation of the distribution ($L_{MC}^{(1)} \pm \Delta L_{MC}^{(1)}$) are presented in comparison with the mean obtained from the entire printed area ($L_{mean}^{(1)}$).

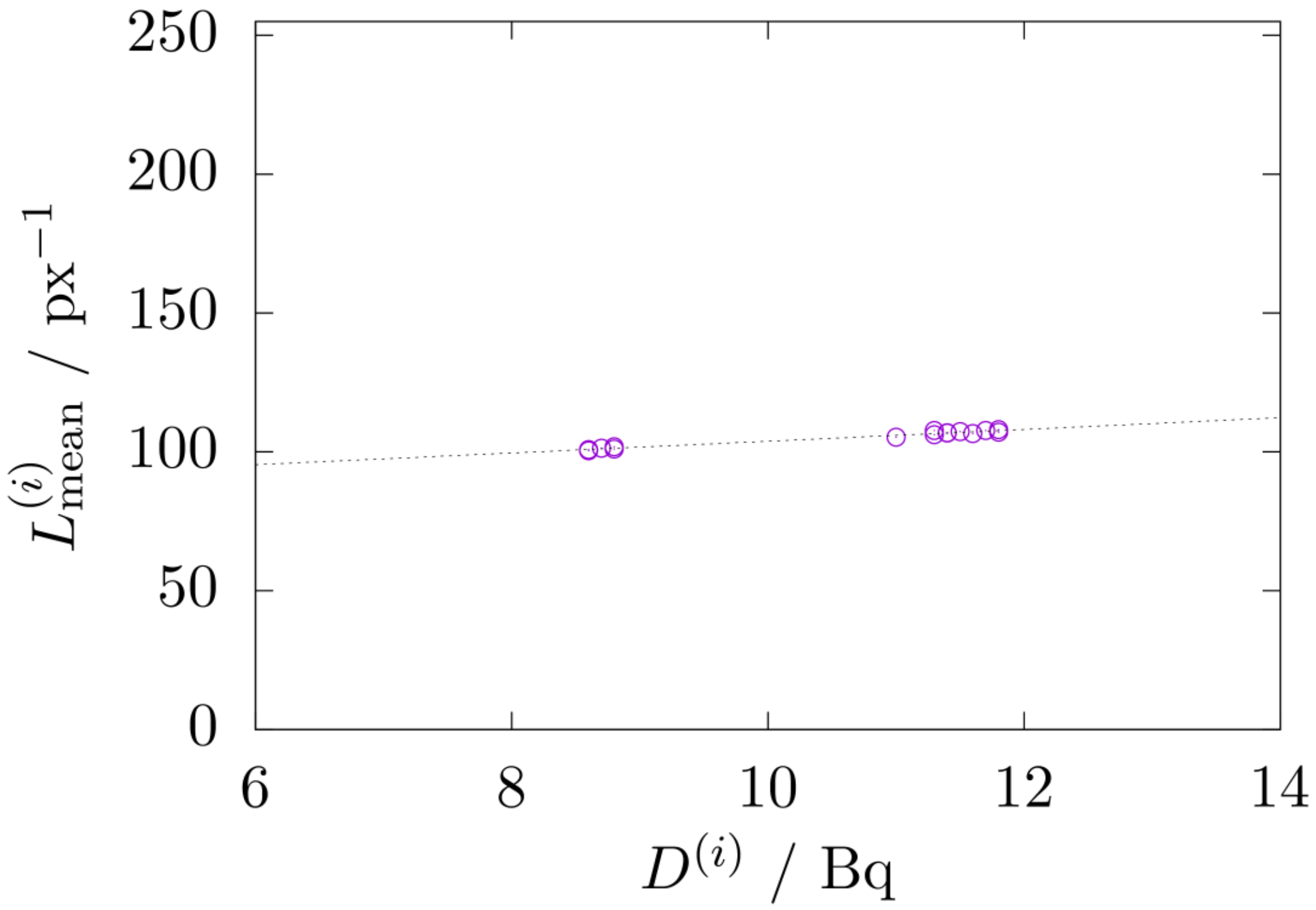

Fig. B.5. Linear relationship between $^{85}Sr$ radioactivity and mean PSL intensity per pixel among 15 $^{nat}Sr$ targets.

Table B.1. Relative uncertainties of $^{nat}Sr$ target thicknesses derived from Monte Carlo analysis

| $^{nat}Sr$ target ($i$) | $\phi$ 1 mm$^a$ | $\phi$ 2 mm$^a$ | $\phi$ 3 mm$^a$ | Gaussian$^a$ | $\delta^b$ |
|---|---|---|---|---|---|
| 1 | +3.1 ± 3.2 | +3.2 ± 1.4 | +3.3 ± 0.8 | +3.3 ± 1.0 | 3.4 |
| 2 | +2.1 ± 3.6 | +2.4 ± 2.2 | +2.3 ± 1.6 | +2.3 ± 1.7 | 2.9 |
| 3 | +2.7 ± 3.7 | +2.7 ± 1.3 | +2.6 ± 0.9 | +2.6 ± 1.0 | 2.8 |
| 4 | +2.3 ± 4.6 | +2.4 ± 1.7 | +2.5 ± 1.2 | +2.5 ± 1.4 | 2.9 |
| 5 | +1.7 ± 4.5 | +1.6 ± 2.2 | +1.3 ± 1.2 | +1.3 ± 1.4 | 2.0 |
| 6 | +2.5 ± 2.7 | +2.9 ± 1.2 | +3.2 ± 0.6 | +3.2 ± 0.7 | 3.3 |
| 7 | +3.3 ± 4.1 | +3.2 ± 2.5 | +3.1 ± 1.7 | +3.1 ± 1.8 | 3.6 |
| 8 | +2.9 ± 2.9 | +2.9 ± 1.6 | +2.7 ± 0.9 | +2.7 ± 1.1 | 2.9 |
| 9 | +2.2 ± 4.0 | +1.6 ± 2.2 | +1.3 ± 1.3 | +1.4 ± 1.5 | 2.0 |
| 10 | +1.7 ± 3.3 | +1.7 ± 1.4 | +1.6 ± 0.9 | +1.6 ± 1.0 | 1.9 |
| 11 | +0.8 ± 6.3 | +0.7 ± 2.6 | +1.0 ± 1.6 | +0.9 ± 1.9 | 2.1 |
| 12 | +1.7 ± 6.0 | +1.5 ± 3.4 | +1.5 ± 1.6 | +1.5 ± 2.0 | 2.5 |
| 13 | +2.3 ± 4.8 | +2.4 ± 1.8 | +2.4 ± 0.9 | +2.4 ± 1.1 | 2.6 |
| 14 | +0.8 ± 6.8 | +1.6 ± 3.7 | +1.2 ± 2.5 | +1.2 ± 2.8 | 3.0 |
| 15 | +2.1 ± 5.2 | +2.5 ±2.2 | +2.7 ± 1.3 | +2.6 ± 1.5 | 3.0 |
| Representative | | | | | 2.8 |

*a* Given as $\Delta_{\mathrm{sys}} \pm \Delta_{\mathrm{stat}}$ style.
*b* Total uncertainty defined as root sum square of $\Delta_{\mathrm{sys}}$ and $\Delta_{\mathrm{stat}}$.